# RESEARCH

# SketchBio: A Scientist's 3D Interface for Molecular Modeling and Animation

Shawn M Waldon[*], Peter M Thompson, Patrick J Hahn and Russell M Taylor II

**Abstract**

**Background:** Because of the difficulties involved in learning and using 3D modeling and rendering software, many scientists hire programmers or animators to create models and animations. This both slows the discovery process and provides opportunities for miscommunication. Working with multiple collaborators, we developed a set of design goals for a tool that would enable them to directly construct models and animations.

**Results:** We present SketchBio, a tool that incorporates state-of-the-art bimanual interaction and drop shadows to enable rapid construction of molecular structures and animations. It includes three novel features: crystal by example, pose-mode physics, and spring-based layout that accelerate operations common in the formation of molecular models. We present design decisions and their consequences, including cases where iterative design was required to produce effective approaches.

**Conclusions:** The design decisions, novel features, and inclusion of state-of-the-art techniques enabled SketchBio to meet all of its design goals. These features and decisions can be incorporated into existing and new tools to improve their effectiveness.

## Background

In his introduction to the "Cellular and Molecular Data" session at BioVis 2013, Tom Ferrin identified two open challenges for molecular biology software. The first was to "Provide easy-to-use learning tools that can still convey complex structures" and the second was to "Develop easy-to-use interfaces that permit facile control of models". We're developing SketchBio to help scientists think about 3D molecular structures and interactions and to communicate them to others.

We found ourselves repeatedly using 2D hand-drawings of complex 3D structures and their interactions in discussions with our close collaborators in cell biology, pathology, and chemistry, despite the fact that the 3D crystal structures of the proteins making up these structures were known. Hanging David Goodsell's excellent "Molecular Machinery" poster [1] showing renderings of many of these molecules and their interactions helped to grow our shared understanding of the structures in the mitotic spindle ( a structure that separates chromosomes during cell division), but real progress was made when we hired an artist to draw 3D scale models of the structures each week and then develop 3D computer models [2].

Our group is not alone. Discussions among collaborators are often done using 2D whiteboard sketches. Presentations often consist of pasted images and 2D Powerpoint animations.

Because of the difficulties involved in learning and using 3D modeling and rendering software, many scientists hire professional computer programmers and/or animators to work with them to create models and animations rather than use these programs themselves. This indirection both slows the discovery process and provides opportunities for miscommunication. As toolsmiths, our aim is to provide scientists with a tool that is so rapid to learn and powerful to use that they can create these models and animations themselves.

We aim to produce a general tool that is widely useful. Many researchers studying cell structure and physiology seek to construct and evaluate dynamic models that incorporate random thermal motion as well as conformational changes induced through intermolecular interactions. Discovering, testing, and communicating hypotheses about these interactions requires the development of complex animated 3D molecular structures. Modeling, simulation, and rendering of these hypothetical scenarios involves using a number of tools

[*]Correspondence: swaldon@cs.unc.edu
University of North Carolina at Chapel Hill, 27599 Chapel Hill, NC, US
Full list of author information is available at the end of the article and databases (PDB, Pymol, Blender, NAMD, etc.) and then converting files to pass geometry and animations between tools. It also involves manual placement and orientation of 3D objects, which is currently done using clunky 2D input devices and by-hand detection and avoidance of collisions. As a result, it often takes a team months to produce an acceptable model or animation. We aim to produce a tool that reduces this to a single person working for hours or days.

This paper describes that tool, SketchBio.

Driving Problems

Fred Brooks points out that the best way to construct a tool that is generally usable is to focus on several very different specific problems and build a tool that solves them [3]. We followed this approach.

The first driving problem for this project is collaborator Susan Lord's desire to construct a protofibril model based on geometric constraints among a set of individual fibrinogen molecules. The protein fibrinogen is the main component of blood clots, where it is converted into fibrin and links together with other fibrin molecules to form strands. Two of these strands join together to form a protofibril, which form thick fibers that make up a large portion of the blood clot. Based on the crystallized structures of fibrin monomers from different species and on only two sets of known interactions [4], Lord sought to construct 3D protofibril structures matching those seen in her data, which suggested a structure in which two fibrin strands twist around each other, and wanted to create a model that shows this interaction at the molecular level. Over several months, she and her students worked with resource staff scientist Joe Hsiao to use the powerful UCSF Chimera tool to construct such a model [5]. Building this model required repeated iteration of hand-placement of two molecules (using multiple 2D mouse interactions), followed by using replication tools to develop candidate models, which were then evaluated against the data. Lord's desired use of SketchBio was to construct this protofibril rapidly and semi-automatically by specifying which location on each fibrin should be in close contact with other molecules and by specifying that the molecules do not overlap. This same capability will enable generation of other self-symmetric structures such as actin filaments and microtubules.

Our second driving problem comes from Peter Thompson in Sharon Campbell's lab. Peter wants to construct 3D models and animations of the interaction between actin filaments and vinculin. Actin filaments are one of the three main components of a cell's cytoskeleton, and the protein vinculin binds to actin filaments, connecting them to other actin filaments or different proteins.

The third driving problem comes from Kerry Bloom, who is constructing models of the mitotic spindle, a structure that separates chromosomes during cell division. As in the Lord case, each step of model generation has required support from an artist, animator, and/or programmer to convert Bloom's concepts and those of his students into geometry for rendering and simulation.

Our final driving problem also comes from the Bloom collaboration. Many proteins beyond cohesin and condensin contribute to mitosis, and Kerry Bloom is interested in them all. His lab is able to fluorescently label both these proteins and chromosome locations and determine relative distances and orientations between pairs of proteins. With accurate localization and tracking for 3D images, these techniques provide partial information on the 3D layout of proteins and chromosomes in wild-type and mutant mitotic spindles. Building models to match this information requires the development of semi-automatic layout of proteins. This will provide a partial set of constraints for the scientist to construct protein-protein and protein-chromosome complexes that match experimental data. With these enhancements, SketchBio will be widely useful to other researchers for the generation of hypothetical protein-complex structures from partial data.

Design Goals

The application-specific needs from our several collaborators can be summarized as a set of domain-independent design goals for SketchBio:

- **Easy to learn and to use.** Scientists must be able to rapidly construct models and animations on their own using interfaces that enable them to concentrate their mental efforts on the design challenge rather than deciphering the interface.
- **Support molecular operations.** It must be easy to load molecules, extract the relevant substructures, describe conformational changes, group molecules, and color according to standard data.
- **Appropriately constrain layout.** Some molecular structures should not overlap, others (drug vs. protein) overlap as part of their function, others (fibrin, actin) assemble into repeated structures. In some cases, the distances between individual elements is known but their 3D layout is not. Supporting all of these cases will enable a biologist to most rapidly explore the space of possible conformations to produce consistent models.
- **Support rapidly iterated, in-context design.** Whether working alone or in a group, understanding the interactions between dozens of molecules requires repeated adjustment of proposed locations and motions. The reasonableness of interactions depends on nearby molecules, which change over time. We have found that generating consistent models requires trying and optimizing many potential solutions before the final model is found.
- **Support high-quality rendering.** Once a proposed model has been completed, static and animated images that use the most-effective lighting and surface rendering techniques are critical to conveying the model and its behavior to others.

Prior Work

*Molecular modeling:* There are many excellent molecular modeling applications that have been extended to include some aspects of high-quality rendering and animation. UCSF Chimera [6], Pymol [7], Graphite Life Explorer[8], and Visual Molecular Dynamics (VMD) [9] are the most relevant. Other software such as Protein Explorer [10] and EZ-Viz [11] (an interface for Pymol) attempt to offer easy-to-use interfaces for exploring molecular structures.

VMD includes direct force-feedback-based placement and manipulation of molecules in the context of driving molecular dynamics. SketchBio provides bimanual control of much larger sets of molecules by reducing the physics to only what is necessary to avoid improper collisions and provide appropriate spacing, enabling large-scale geometric modeling and animation. GraphiteLifeExplorer includes the ability to position and twist segments of DNA and interpolate the sections between them (its DNA modeling tools go beyond what is available in SketchBio), but does not yet perform collision detection between molecules, the ability to support animation, or the ability to maintain specified distances between objects needed by our collaborators.

We considered the approach of extending the interaction and rendering capabilities of one of these tools, but this would require re-implementing existing rendering techniques and continual updating as new rendering advances are made. We instead decided to harness the power of the existing tools through their built-in scripting languages (SketchBio has used both Pymol and Chimera to load, surface, select, and label molecules by partial charge and other inputs).

*Rendering:* There are also excellent general-purpose rendering programs (such as the commercial Maya and open-source Blender applications) and microscope-simulation rendering tools (such as UNC's Microscope Simulator [12]). Several groups are building molecule-specific loaders that plug into these programs, such as Autofill/Autopack/Autocell [13], and Molecular Maya [14]. The Bioblender package also leverages Blender for molecular modeling and supports collision detection [15]. These each require the scientist to learn the underlying complex rendering tool plus additional plug-in interfaces, making them less easy to learn and use. None of these tools currently support constrained layout along with rapidly-iterated, in-context design.

Molecular Flipbook [16] aims at similar goals to SketchBio, providing an easy to use molecule-focused real-time interaction environment coupled to offline rendering using Blender and FFMPEG. It does not currently support constraints on layout or bimanual interaction for rapid 6-degree-of-freedom placement. We also considered this approach, but providing full capability would require re-implementing many existing capabilities already available in molecular modeling tools and tracking new features as they are developed.

*Interactive Animation:* The Molecular Control Toolkit [17] is also aimed at molecular modeling, providing gesture- and speech-based user interface primitives to control motions of molecules with a Kinect or Leap Motion device [17]; it provides an API that can be used to connect their controls to existing molecular modeling applications. These do not by themselves meet the needs of our collaborators, but could be used within SketchBio as a separate front-end interaction interface. SketchBio uses similar two-handed 6-degree-of-freedom input devices (the Razer Hydra or two WiiMote controllers), adding collision detection and several custom capabilities, and tying the resulting system into existing powerful molecular modeling and rendering tools to produce a complete system for thinking, modeling, and rendering.

Another tool aimed at simplifying the creation of molecular animations, PresentaBALL[18], uses an interactive web interface to an existing molecular modeling tool [18]. This allows widespread use by non-experts to develop presentation materials for training. SketchBio provides a custom interface for experts to use as a thinking aid that is tied to a powerful rendering engine to produce animations.

SketchBio's bimanual 6-degree-of-freedom manipulation sets it apart from all of the applications described above because it lets the user move molecules and craft animations more rapidly and with less mental effort than tools that use a mouse and keyboard to manipulate objects. Its support of appropriately-constrained layout using several features (configurable collision detection, spring-based layout, and crystal by example) meet needs expressed by our collaborators that are unmet by any published tool.

*Interactive Rendering:* A common bottleneck in interactive modeling and animation applications is the speed of rendering a complex scene. Sketchbio requires real-time rendering due to the nature of its input – objects on the screen must be move with the user's hand as if the user were actually holding them.

One approach to improving rendering speed is to reduce the complexity of the objects that are drawn. This is done by replacing objects with imposters which have simpler geometry. One type of imposter is a simplified version of the geometry that is textured to look like the more complex version [19, 20, 21]. Another common imposter is a square that has a pre-rendered image of the more complex object as its texture. As long as the viewpoint stays near the same position, discrepancies between the imposter and the actual geometry are be small [22, 23].

The level of simplification of an object can also be dynamically determined according to the amount of the allotted rendering time that is needed to draw each level of detail.

Another approach to enabling interactive rendering of complex design spaces is to precompute an ensemble of possible solutions and then interactively explore the design space by directly manipulating portions of it and morphing between existing solutions [24]. The space of potential molecular interactions for dozens of molecules is so large, and the ease of testing and rendering each configuration so small, that it was more efficient for SketchBio to directly model and render.

SketchBio uses Chimera and Blender to simplify geometry and the Visualization ToolKit (VTK) library to adjust rendered level of detail [25].

*Collision Detection:* In many models and animations, molecules should not overlap one another. If there are $n$ molecules in the scene, then each pair of molecules must be tested for collision. This has a complexity of $\mathcal{O}(n^2)$ in the number of molecules. However, there are typically far fewer collisions than potential collisions and so optimizations can reduce the expected complexity. The best expected complexity uses sweep and prune methods and assumes the primitives are sorted along one dimension. This is $\mathcal{O}(n + c)$ where c is the number of colliding pairs [26].

Another approach uses space partitioning to rule out unnecessary tests. The PQP library from the UNC GAMMA group uses a bounding volume hierarchy [27]. An alternate is to divide space into bins. Only primitives in nearby bins need to be tested. This type of algorithm is especially effective on GPUs where many local groups may be run in parallel [28].

SketchBio directly links to PQP and uses it for basic collision detection. It extends these techniques in ways that are specific to the kinds of molecular models being formed to gain an additional order of magnitude reduction in collision tests for some objects.

*Mash-ups:* Individual capabilities of web-based applications such as Google maps and real-estate listing databasese have been effectively combined to provide combination tools that include the best parts of each. A system for seamless integration of applications for visualization was done by Rungta et. al by adding a layer above all of the applications of interest to pass events back and forth [29]. SketchBio takes a similar approach, using a novel core component that provides interactivity and custom features but using scripting interfaces to harness the significant modeling and rendering capabilities of existing tools into a seamless workflow.

## Methods

SketchBio is a system for understanding subcellular biology through the building of complex 3D macromolecular structures and animating the structures over time.

The modeling and rendering of these hypothetical structures currently involves using a number of tools and databases and converting files and data to pass between tools.

SketchBio harnesses state-of-the-art tools and libraries into a seamless workflow. It brings best-practice interaction and display techniques to bear on molecular modeling, including bimanual real-time direct interaction and shadow-plane depth cues. It adds three novel features that accelerate this workflow: crystal-by-example, pose-mode physics, and spring-based connectors. Its design decisions (a direct-manipulation, real-time interface; harnessing tools rather than re-implementing techniques; and making a system usable in the scientists' labs) led to a system that met all of the design goals.

### System Overview

Figure 1 shows a screenshot of the SketchBio user interface with a group of three actin molecules (left) and the tail region of a vinculin molecule (right). SketchBio uses imposters with simplified surface geometry while developing the animation, but uses full resolution models for final rendering. The small white spheres follow the two tracked hand-held controllers. Status information is shown in the lower left. The current animation time point is shown in the lower right.

Molecules in SketchBio are represented as rigid surfaces approximating the Connolly Solvent-Excluded Surface of the molecule. These were chosen because our collaborators used surface models in their current work for all four driving problems. The surfaces may use solid colors, be colored by surface charge, or be colored by their nearness along the protein backbone to the N-terminus or C-terminus of the protein. Chimera is used to calculate and export the datasets for the latter two coloring schemes using PDB data.

Object selection is indicated by drawing the outline of the oriented bounding box of the selected molecule molecules. Color of this outline indicates whether a group or single object is selected. (An earlier design showed the selected object in wireframe, but this was found to disrupt perception of the orientation of the molecule.)

A set of "gift ribbons" drawn on the oriented bounding box indicates that an object has a keyframe at the current time. This was chosen to minimally obscure the molecule and selection indicators.

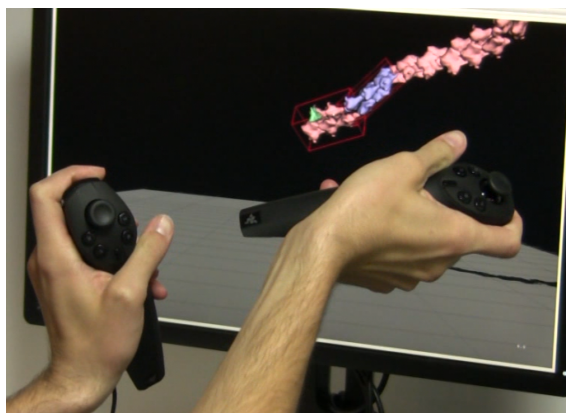

**Figure 2** The left hand sets the base molecule while the right hand positions the copies in this two-handed construction of an actin fiber.

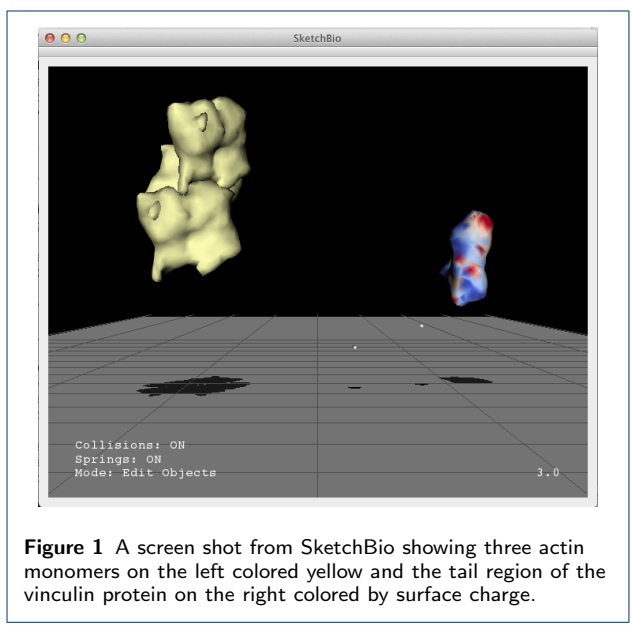

**Figure 1** A screen shot from SketchBio showing three actin monomers on the left colored yellow and the tail region of the vinculin protein on the right colored by surface charge.

### State-of-the-art Capabilities

*Bimanual Interaction:* Bill Buxton and others have described the benefits of two-handed (bimanual) interaction. He and others observed that bimanual manipulation brings "two types of advantages to human-computer interaction: manual and cognitive. Manual benefits come from increased time-motion efficiency, due to the twice as many degrees of freedom simultaneously available to the user. Cognitive benefits arise as a result of reducing the load of mentally composing and visualizing the task at an unnaturally low level imposed by traditional unimanual techniques" [30].

As seen in Figure 2, SketchBio brings bimanual interaction to the construction of macromolecular structures. The entire interface is built around a set of world and root-object manipulation controls in the non-dominant hand and a set of individual-element manipulation controls using the dominant hand.

SketchBio uses a pair of Razer Hydra controllers to provide two 6-DOF trackers, each of which also has several buttons, a hi-hat controller, and an analog input. This enables a very expressive set of verbs (buttons), nouns (selection via 3-DOF positioning), and adjectives (magnitude via analog inputs, viewpoint via hi-hat, and pose via a combined 12-DOF tracking). This avoids the need for the system to recognize a large set of ambiguous gestures, as is the case for video-based user input. Use of this device enables the interface for moving objects to mirror a task users are already familiar with, namely reaching out, grabbing an object and moving it to a new position and orientation.

Using one of the buttons to switch between modes provides a sufficiently-large space of commands that almost all operations can be performed without putting down the controllers. The keyboard and mouse are used to name proteins and files on initial loading, and to set precise values as needed for one or two operations.

*Shadow Plane:* Because selection in SketchBio requires placing the tracker within the bounding box of the object, determining the relative depth between tracker and object is an important and often-performed task. Initial testing of the application revealed that determining the relative depth between an object and the tracker or between two objects was the most difficult part of using SketchBio. Because widespread adoption would be limited by requiring stereo displays and head tracking, we sought another solution.

Hendrix and Barfield found the most effective techniques for aiding in depth estimation are a textured plane and lines dropped from the center of an object to the textured plane [31]. To provide additional depth cues, SketchBio displays a ground plane that is always rendered below the viewpoint no matter the direction or position of the viewpoint and projects the shadows of objects onto this plane. The trackers also cast shadows onto this plane (which are darker and larger to highlight them). SketchBio assumes a light infinitely far away in the default camera's up direction which gives the same absolute position against the textured surface as the drop-lines while also giving information about how close the boundaries of two objects are to each other. The user can also rotate the camera while leaving the light and shadow plane fixed to get a better understanding of the scene through motion parallax [See Figure 3].

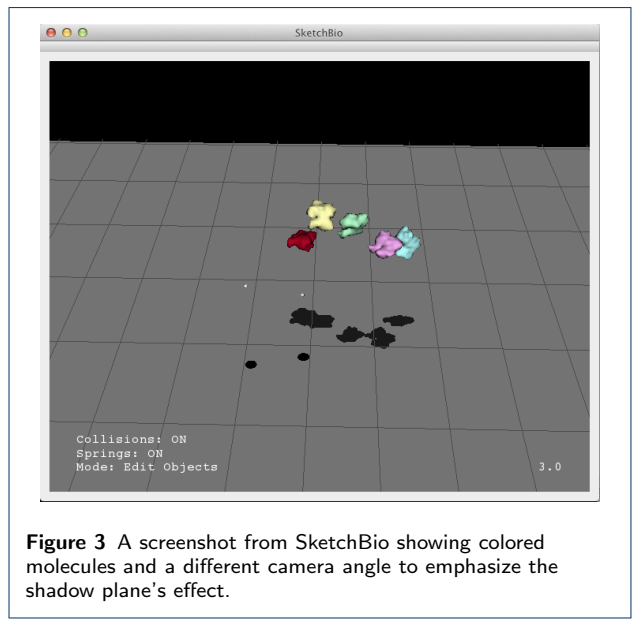

**Figure 3** A screenshot from SketchBio showing colored molecules and a different camera angle to emphasize the shadow plane's effect.

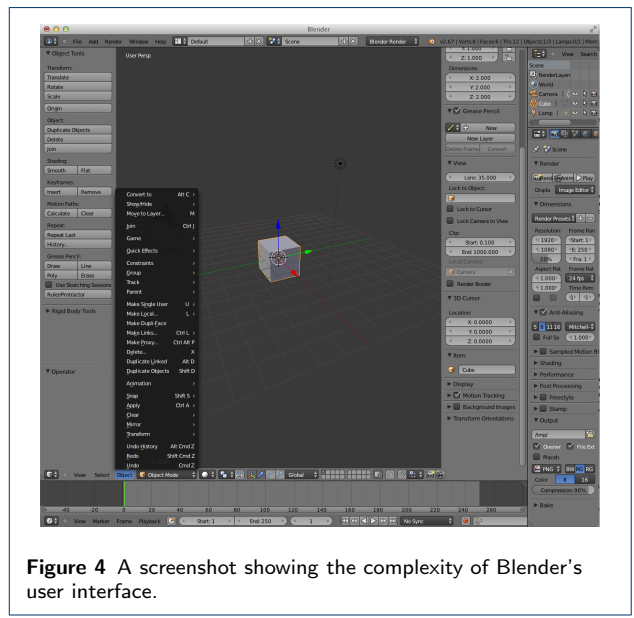

**Figure 4** A screenshot showing the complexity of Blender's user interface.

*Animations:* For scientists creating animations of molecules, SketchBio provides a basic interface to a much more complex system. Blender is a production level animation and rendering tool that has an extremely complex user interface with dozens of hotkeys, menus and buttons (see Figure 4). Blender also has a Python scripting interface that provides access to all of its functionality. SketchBio uses this scripting interface to create its animations and render them in a high quality rendering engine, but provides a much simpler user interface. SketchBio provides a set simple operations that is sufficient to meet the animation needs of our driving problems: moving along the video timeline, setting keyframes on objects and viewing a low resolution animation preview.

Keyframes can modify color and grouping information as well as object position and orientation. These values are interpolated between keyframes using splines to produce smooth motion and changes. The effects of this interpolation can be easily seen by the user by moving along the timeline or using the built-in animation preview. The scene is exported to Blender with a set of predefined global settings for effects and position of light sources to produce a full-quality rendering.

*Grouping:* Grouping of molecules eases construction of larger order structures and provides smooth animation of objects that should moving together without the small variations that even the most careful hand placement causes. Copy and paste is also implemented (both single objects and groups can be copied and pasted) even between sessions. Additionally, a group of molecules constituting a structure that a user wants to use multiple times in different projects can be saved and then imported, eliminating the need to rebuild large structures. Molecules can be added to groups or removed from them at keyframes.

*Importing Molecules:* SketchBio generates molecular surfaces using UCSF Chimera via Python scripting. A custom plugin (ExportVTK) was written for Chimera's Python interface to export additional data from Chimera in the VTK file format. This plugin was contributed back to the Chimera developers and is now part of the standard source distribution. This data includes residue and chain identifier that map to a specific location on the surface and electrostatic potential on the surface. SketchBio can use these data sets to color the objects (see figure 1).

Novel Capabilities

To meet the needs of our collaborators, SketchBio supports novel operations beyond those available in the programs and libraries that it harnesses. These include "pose-mode physics" that enables rapid docking of one protein with others, a "crystal by example" mode that enables rapid formation of polymer molecular chains, and spring-like connectors to maintain expected distances between molecules. Each of these is described, along with how they enable optimization of collision detection.

*Pose Mode Physics:* Object motion in SketchBio is accomplished by applying forces and torques to pull towards the tracker location and orientation. This can result in the object lagging behind but also smoothes motion, especially rotation.

Standard rigid-body dynamics was used as the original collision response in SketchBio. Because the manipulated object pushed other objects around, this caused difficulty in assembling molecular groupings.

This was solved by introducing "pose-mode physics", where the only objects that move are those directly being manipulated. other objects do not move when collision response forces are applied. This also greatly reduces the time taken to compute collision detection (as described later).

The first implementation of pose-mode physics onle moved the object if its new location after being pulled by the tracker-attracting forces would be collision free. This caused objects to become stuck together and difficult to pull apart because tracker rotation usually introduced collisions even as they forces pulled objects apart. This also prevented sliding objects along each other, which scientists often wanted to be able to do.

In the final implementation, where collision response forces act on the object being manipulated, objects can be slid along one another but not collide.

*Crystal-by-example:* Repeated structures that formed by replicating a single protein are common in biology (actin, microtubules, fibrin, etc.), so the "crystal-by-example" feature was added to support their construction. Our collaborators wanted to construct variants of such structures to study the changes caused by mutant proteins and to understand their native packing for comparison to electron microscopy images.

A similar problem is addressed in [8] for DNA molecules by letting users edit placement and twist of selected base pairs and interpolating in between these. That system forces the resulting structure to follow a specified path. Crystal-by-example inverts this to show the structure resulting from a specified packing geometry: the user places two molecules relative to one another in six degrees of freedom and SketchBio repeatedly applies same transformation for other copies to generate chains of molecules. Each replication of the base molecule follows the example set by the first two molecules, with the third molecule's placement relative to the second molecule being the same as that of the second molecule to the first, and so on. In this way, a repeated structure is formed by manipulating only one pair of molecules rather than tediously moving each individual piece to its proper place, speeding up the process of building structures.

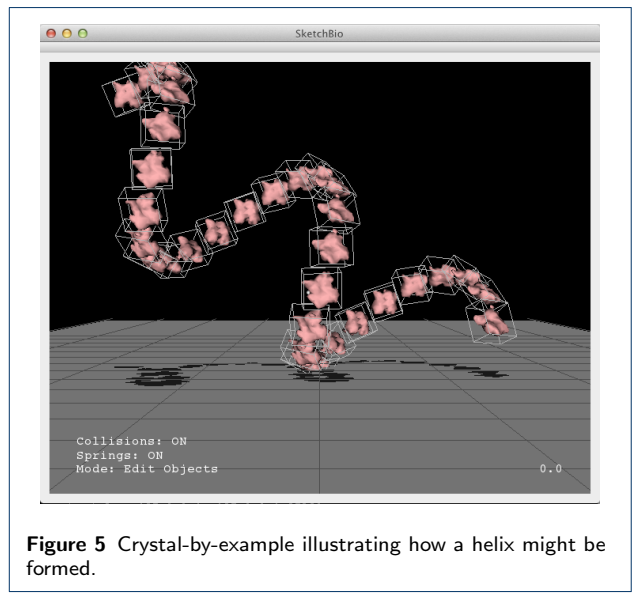

**Figure 5** Crystal-by-example illustrating how a helix might be formed.

This feature uses two copies of a molecule (A and B) to define an entire repeated structure. Given $T_A$ and $T_B$, the transformation matrices that define the positions of A and B relative to the world origin, the transformation from A's coordinate system to B's coordinate system, $T_{AB} = T_A^{-1} * T_B$, can be computed.

B's position can be rewritten $T_B = T_A * T_{AB}$. The next repeated molecule, C, has position $T_C = T_B * T_{AB} = T_A * T_{AB}^2$. This can be extended to generate a chain including an arbitrary number of molecules.

Many biological structures including actin fibers and microtubules (major components of a cell's cytoskeleton) form in structures that can be defined this way. Figure 6 shows an actin fiber generated this way in SketchBio. By providing live updates of the entire structure as the initial two objects are manipulated, SketchBio lets the scientist explore potential structures in real time.

The extent to which the user can control fine-grain maniuplations of the molecules depends on the input device in use, because resolution varies by device. Because some structures have a known transformation from one molecule to the next, SketchBio (like other programs) lets the user input the transformation directly.

*Collision Detection in Pose Mode Physics and Crystal-by-example:* In pose mode, collision tests between objects that the user is not interacting with can be skipped because these objects do not move. This means that only collisions involving the objects that the user is moving need to be checked. This reduces the number of collision tests to $m*n$ where m is the number of objects that the user is currently moving. The typical number of objects that the user moves at a time is 1 or a small constant (in the case of moving a group), which reduces the number of collision tests needed to $\mathcal{O}(n)$ in this expected case.

There are two ways that the user can interact with a crystal-by-example structure: moving the entire structure as a unit, or adjusting the internal transformation to change the shape of the structure. In the first case, only collision tests between the structure and the other objects in the scene need to be done, and the above bound applies to the number of tests.

In the second case, the internal structure does change and both internal and external collisions must be tested. External collisions must test every object in the structure with every external object as above.

In internal case we can leverage the known relationship between the objects to perform fewer tests. Let $X_i$ be the ith object in the crystal by example structure with $X_1$ and $X_2$ being the two base object in the structure. Let $T_{i,j}$ be the transformation matrix from $X_i$ to $X_j$. The definition of the crystal-by-example structure is that $T_{i,i+1}$ is the same for all $i$ and the geometries of all the $X_i$s are the same. Because the geometries and transformations are the same, if there is a collision between the *ith* and $(i+1)th$ objects anywhere in the structure, then there is also a collision between the $1st$ and $2nd$ objects. Thus testing only this one pair performs the work of n-1 tests where n is the number of objects in the structure. This same argument holds for any $i$ and $i+k$, the $1st$ and $(k+1)th$ objects have the same relative positions and the same collisions. Thus only the $1st$ object in the structure needs to be tested against the others which allows $\mathcal{O}(n)$ tests to suffice for all internal collisions in a repetitive structure of $n$ elements.

*Connectors:* SketchBio also has connectors that can be added between objects. These can act like springs and apply forces to keep objects positioned relative to each other or they can simply indicate that two objects are connected. Many proteins have regions for which the structure is unknown and these regions can be represented with these connectors. Responding to a scientist's request, the connector end can be snapped to the N-terminus or C-terminus of a protein, removing the difficulty of precise hand placement.

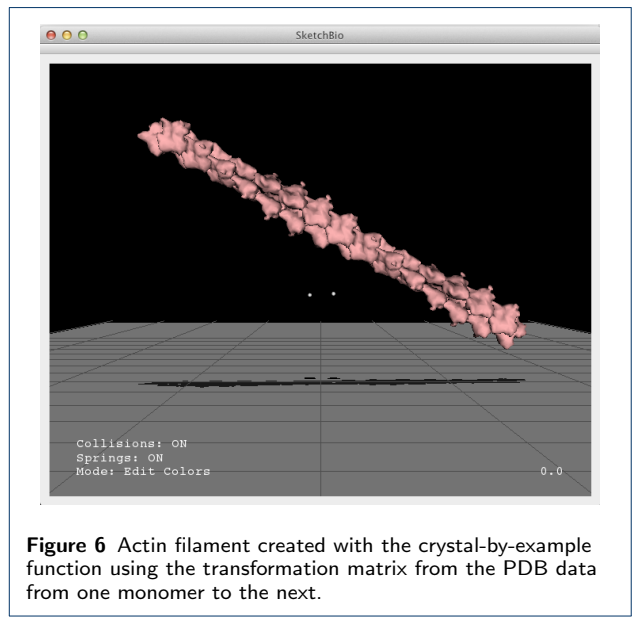

**Figure 6** Actin filament created with the crystal-by-example function using the transformation matrix from the PDB data from one monomer to the next.

When acting as springs, connectors can have non-zero rest length. When editing a set of proteins some of whose separations are known experimentally (through two-color fluorescence labeleing, FRET, or other techniques as in our final driving problem), this can be used to specify soft constraints on the 3D layout of the proteins, guiding the scientist away from impossible structures. This greatly reduces the conformation space that must be searched to determine molecular arrangements.

### Architecture

The architecture of SketchBio is shown in Figure 7. SketchBio harnesses external programs when possible (PyMOL, Chimera, Blender) and uses existing libraries for other core functions (VTK, PQP, VRPN). It maps from dozens of controls in Chimera and hundreds of controls in Blender down to 4 input options and about 20 modeling and animation controls to streamline the tasks needed for creating structures and animations.

Exporting data to Blender is done by writing a script to be run on Blender's Python interface to produce the animation. When exporting to MicroscopeSimulator, SketchBio writes out a Microscope Simulator XML project file and loads the project into MicroscopeSimulator.

Objects can be loaded into SketchBio as .obj files from any program that writes this format or directly through the GUI (via harnessing UCSF Chimera from the PDB or a local .pdb file). Because VTK is used in SketchBio, any file format that VTK can read could be imported with relatively minor changes.

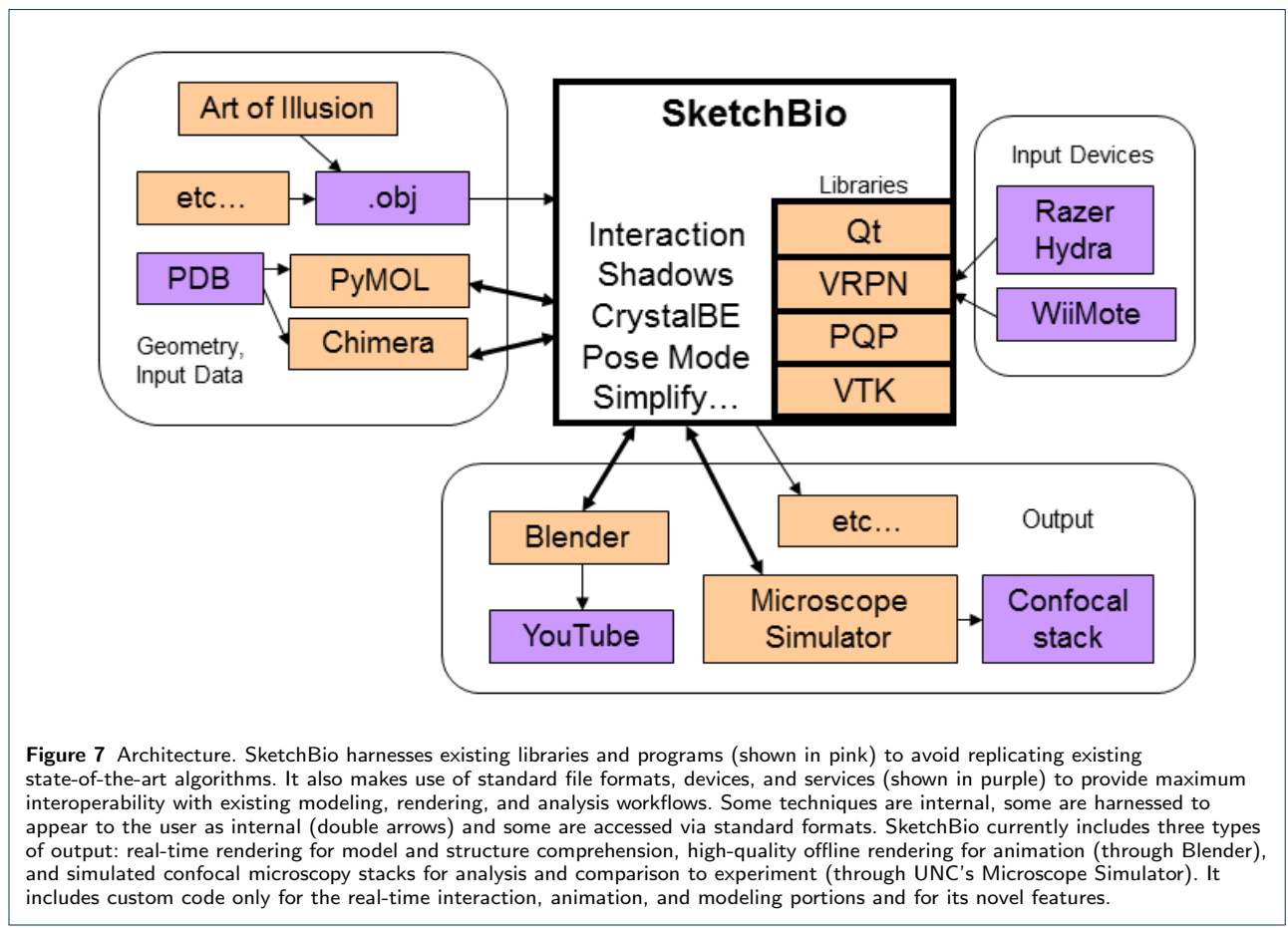

**Figure 7** Architecture. SketchBio harnesses existing libraries and programs (shown in pink) to avoid replicating existing state-of-the-art algorithms. It also makes use of standard file formats, devices, and services (shown in purple) to provide maximum interoperability with existing modeling, rendering, and analysis workflows. Some techniques are internal, some are harnessed to appear to the user as internal (double arrows) and some are accessed via standard formats. SketchBio currently includes three types of output: real-time rendering for model and structure comprehension, high-quality offline rendering for animation (through Blender), and simulated confocal microscopy stacks for analysis and comparison to experiment (through UNC's Microscope Simulator). It includes custom code only for the real-time interaction, animation, and modeling portions and for its novel features.

Design Decisions

We list here some design decisions that helped SketchBio achieve its goals.

*Bimanual, 6-DOF interface:* SketchBio's two-handed interface differs from that of most existing modeling and rendering tools. This has the deficit of taking the user's hands away from the keyboard, which requires them to put down the interaction devices to enter text and specific numerical data. Our collaborators report that this small negative is greatly outweighed by the ability to rapidly perform the more-common and more-challenging tasks of specifying positions, viewpoints, and animations in full 6 degrees of freedom. The ability to move both the world/viewpoint and an animated molecule enables rapid planning of scenes and the ability to simultaneously manipulate both of the molecules that are coming together in an interaction are two examples of what is enabled.

The workflow tends to stratify: initial loading of the kinds of molecules to be used in an animation happens first (with keyboard and mouse). Then positions, viewpoints, and animation are described using the buttons and controls on the two hand-held controllers. Finally, saving the file and rendering are again performed with the keyboard and mouse. The use of rich input devices enables the bulk of the action to take place from within the 3D environment, accelerating the most-challenging parts of model and animation development.

*Harness, don't re-implement:* The design of SketchBio avoids reimplementing existing features where possible, instead using Python scripting to control subprocesses to perform these operations. When reading in PDB files, instead of writing a PDB file reader, SketchBio calls UCSF Chimera as a subprocess to read in the protein and create a displayable surface from it. Instead of writing a new rendering library, SketchBio uses the Python scripting interface of Blender to create a Blender project that will produce the desired animation. SketchBio uses the open source Qt and VTK[25] libraries for its user interface and internal rendering and the open source Proximity Query Package (PQP) for collision detection [27]. The VRPN library is used to communicate with the Razer Hydra input device [32].

One significant risk encountered when harnessing existing programs is that future versions of the programs will not support required features, or will require modifications to the harness. This can make maintainence challenging. To address this, each SketchBio release includes a list of specific versions of the wrapped programs which with it is known to be compatible and have selected programs to harness that continue to make old versions available (Chimera still releases installers from 2002 and Blender from 2003). It also includes copies of custom plug-ins and scripts that are not yet part of the harnessed packages' released versions.

Another risk is that the packages used will not be obtainable in the future, or for an operating system of interest. SketchBio has been able to mitigate this risk by selecting open-source programs to harness.

To measure the re-use of functionality, we compare (1) the number of state-of-the-art operations harnessed from existing tools: Chimera (connecting to the protein data bank, parsing PDB file, selecting subunits, generating surfaces, generating data sets on the surfaces, simplifying surfaces), Blender (surface rendering, directional illumination, transparency, ambient occlusion, parallel rendering, frame storage), and Microscope Simulator (point-spread-function 3D blurring, TIFF stack generation) and (2) the number of internally-used existing libraries: VRPN (reading from general peripheral devices), PQP (multi-object collision detection), VTK (geometric operations, real-time rendering, level-of-detail rendering, object positioning, spline interpolation) to (3) the number of custom operations (crystal by example, pose-mode physics, drop shadows, bimanual interaction modes, spring connectors, grouping and animation). We find that most of the operations are supported by existing tools. We compare this against other tools we have built to support biomedical applications [33]. SketchBio has a much better re-use ratio than tools which similarly span different domains (nanoManipulator, Camera Calibration, Chromatin Cutter, Template-Based Matching) and is on par with tools that are basically wrappers for calls to a single library (ImageTracker, Microscope Simulator). It bests several of our single-domain tools (Video Spot Tracker, Video Optimizer, and WebSlinger). Furthermore, the scripting interfaces enable rapid inclusion of additional features from external programs without re-implementation.

*Usable in-house:* We have in the past built high-performance molecular graphics applications for scientists that used head-tracked stereo, wide-area tracking systems, and force-feedback displays [34, 35, 36, 37, 38, 39]. The scientists who were willing to travel to our laboratories to use them received great benefit, but we wanted SketchBio to be more broadly available. To maximize its impact, SketchBio is designed to run on a laptop or desktop system such as a scientist would have at home or in their laboratory and to use inexpensive commercial input devices.

## Results and Discussion

SketchBio has been used by a several scientists and has demonstrated success in meeting its design goals.

*Easy to learn and use:* To measure the ability of scientists to learn and use the system, we showed SketchBio to a visiting graduate student from NIH. She is interested in using the system to study the proteins involved in cell focal adhesions. After a 30-minute training session where she saw us using the system, she was able to use SketchBio to load, replicate, and place the molecules into relevant configurations.

After similar initial training, and with access to the manual, biochemistry graduate student Peter Thomson used the system to generate both static and animated multi-protein models.

Peter created a model to compare the importance of electrostatics between two different models for vinculin tail interaction with actin [40][41].

He also created an animation of vinculin binding to an actin fiber for use in a talk, based on the model presented in [42]. This video used crystal-by-example to generate the actin and used traslucent connectors to indicate the connection between the head and tail domains of vinculin – a region for which there is no crystal structure. The model in SketchBio is shown in Figure 8 and a frame from the resulting video at approximately the same time is shown in Figure 9.

Peter produced both a SketchBio animation and a Microsoft Powerpoint animation of molecules (using images of molecules pre-rendered from a single viewpoint), as shown in figure 10, to test their relative speed and effectiveness. The Powerpoint animation took 50 minutes of concentrated effort to produce, while the SketchBio animation took 100. He reports that the Powerpoint animation failed to accurately show rotation of the vinculin tail domain, to show the linker region that scales as the domains move apart, to show a change in actin movement rate, and to accurately portray relative size and orientation of the molecules. He reports that the increase in correct presentation of the science was was well worth the increased time.

*Support rapidly-iterated, in-context design:* To measure the speed of complex model construction, we repeated a task using SketchBio that had been done beforehand. Constructing the protofibril models for Susan Lord took resource computer scientist Joe Hsiao 3-3.5 hours by hand-editing transformations within Chimera (a task challenging to teach to biologists). Using an early prototype of SketchBio, he constructed the protofibril seen in Figure 11 in 1.5 hours (a task we'd expect a collaborator to do just as rapidly). The lack of depth cues became apparent as Joe spent most of the time trying to figure out the relative depth between the tracker and the molecules, prompting us to add the shadow plane. With this addition and other features, Joe reconstructed the model in 35 minutes. In all cases, the desired model was known a-priori; all cases measure time on task and do not count the time spent learning how to use the tool. In this case, SketchBio enabled model creation in about one-fifth of the time for a case of interest to our collaborators.

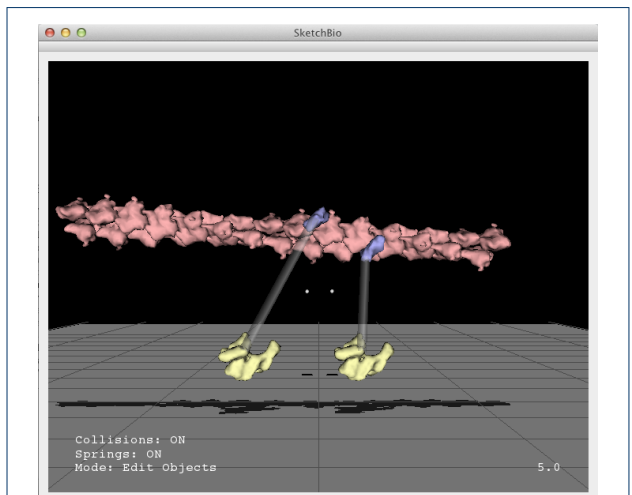

**Figure 8** A scene from a video created by Peter Thompson in SketchBio. Approximately the same timestep is shown rendered at its full resolution in Figure 9

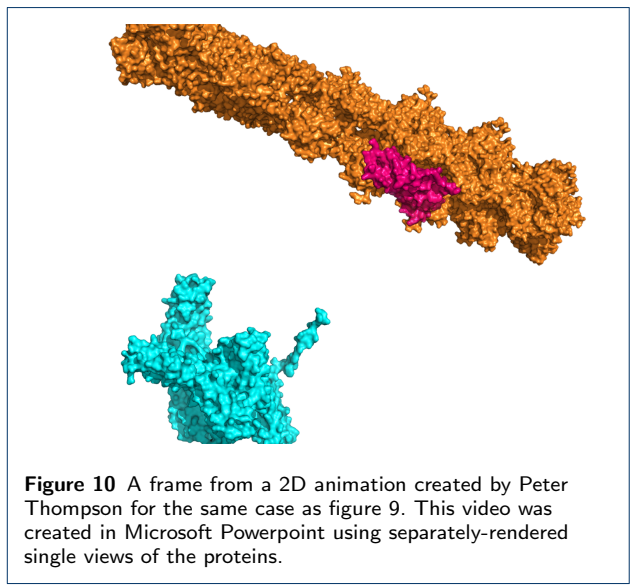

**Figure 10** A frame from a 2D animation created by Peter Thompson for the same case as figure 9. This video was created in Microsoft Powerpoint using separately-rendered single views of the proteins.

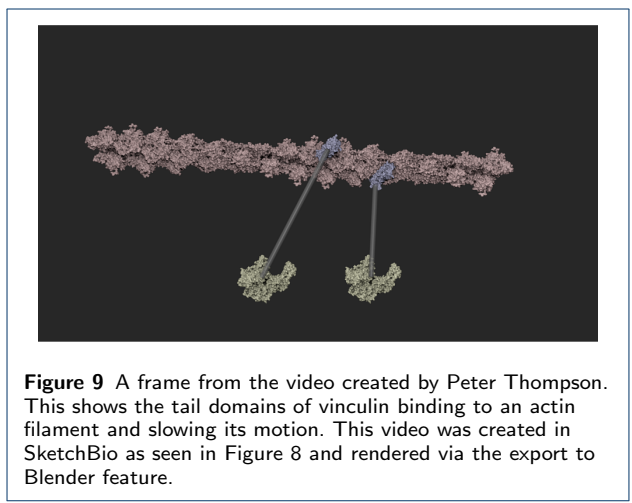

**Figure 9** A frame from the video created by Peter Thompson. This shows the tail domains of vinculin binding to an actin filament and slowing its motion. This video was created in SketchBio as seen in Figure 8 and rendered via the export to Blender feature.

To further measure the effectiveness of SketchBio for the rapid construction of animations, we used it to create an animation of actin and vinculin (see supplemental materials). We were able to load the molecules, replicate them, place them, plan camera and motion paths, and start rendering in half an hour. The first-person design view and available pre-animation were crucial to this process, enabling design intent to be rapidly translated into action and evaluation, resulting in uninterrupted planning and design iteration.

These cases indicate that a series of brief training videos plus the online manual should suffice to get new users started, that scientists are able to use SketchBio on their own, and that SketchBio compares favorably to existing methods of producing animations and structural models.

*Support molecular operations:* The video in supplemental materials shows that a user who is familiar with both tools is able to load, select subsets, and attach two molecules six times as fast using SketchBio as using the combination of Chimera and Blender. As part of development, we created a Chimera plug-in that exports the standard molecular labelings (main-chain index, partial charge, etc.) in a VTK data structure, enabling them to be used to color the molecules. Scientist are able to use familiar PDB file and substructure names to load and extract subsets of molecules. The animation and object-grouping features have been used by our collaborators to produce models and animations meeting their needs.

*Appropriately constrain layout:* Pose-mode physics, with the option to turn it off, supports both preventing and allowing overlap between molecules, as appropriate to the task. The crystal-by-example feature has been used to produce both protofibrils and actin filaments from their monomers. Fixed-length springs provide the ability to rearrange sets of moluecules while maintaining the specified pairwise separations among them.

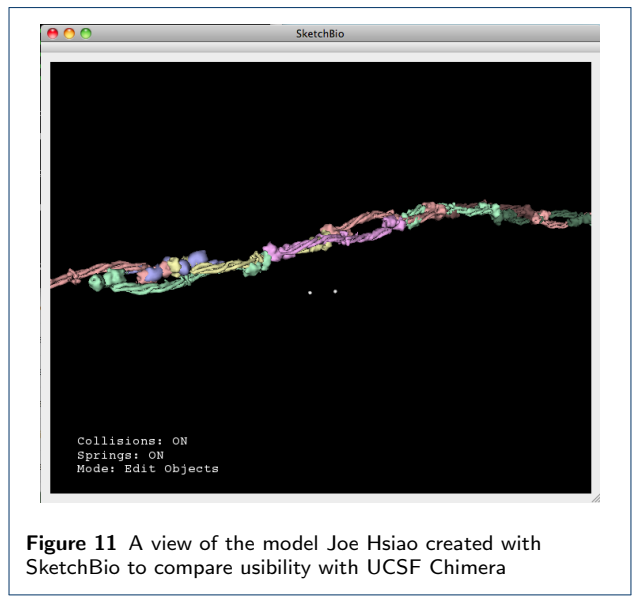

**Figure 11** A view of the model Joe Hsiao created with SketchBio to compare usibility with UCSF Chimera

*Support high-quality rendering:* The image frames in the paper and video in the associated online materials demonstrate full-resolution rendering with intra- and inter-object shadowing displaying both complex local shape and 3D relationships between objects.

## Limitations and future work

After successfully using SketchBio for his initial needs, Peter has requested new features. He is particularly interested in using SketchBio as a thinking tool to determine how mutations in vinculin turn normally-straight actin filament bundles into helices. Forming a model to fit experimental data can is challenging. We are working on a module to optimize the placement of molecules based on a set of constraints. The resulting optimization algorithms will enable other scientists to semi-automatically construct multi-protein structures that match negative stain electron microscopy images.

Our collaborators' projects have so far involved hand-placed molecules of density sufficiently small to be understood when all of them are visible. Thus, SketchBio does not yet support automatically-placed molecules to fill the space, nor does it require complex occlusion-handling procedures. As the user base grows, we expect to need to harness importance-based rendering techniques and autofill algorithms to handle a large number of background molecules. These more complex scenes will also require the ability to label important molecules.

The motion of objects could be changed to directly map the user's hand motion instead of moving toward it via force and torque being applied. This would provide a direct mapping of hand location to object location and possibly a better interface. This could be combined with the collision detection type where objects are only allowed to move to a location if the result is collision free. The disadvantage of this approach is that the smoothing by the forces and torques will not occur; transmitting any jitter in the device input directly to object motion.

SketchBio currently supports only rigid structures for modeling and collision detection. For the collaborations we have worked with deformable models were not needed and we felt our time was better spent providing them with features they had requested.

Molecular dynamics simulation is something SketchBio does not do directly. This decision was motivated by the cost of performing the molecular dynamics and the requirement to provide real-time user interaction. SketchBio may eventually harness an external molecular dynamics simulator, but SketchBio will only be used to specify input configurations for the simulation or easily create videos from its output. While SketchBio will not support molecular dynamics directly, a molecular docking capability involving two individual molecules could be added.

To avoid dependence on a particular hardware vendor, SketchBio is being actively ported to use a pair of Nintendo WiiMote controllers instead of the Razer Hydra controller. Its use of the VRPN library supports switching devices by renaming the device and input for each function; a general-purpose mapping layer has been added that reads from a configuration file to enable the user to customize this remapping. This enables new SketchBio users to continue to use the tool until the next-generation Razer Hydra is released.

One consequence of the choice to provide a uniform environment that wraps functions from other programs is that not all features of the wrapped programs are available from within SketchBio. This limitation is mitigated by enabling the user to export Blender files for later offline rendering and to import arbitrary geometry, but then the user has to learn the complexities of the other tools to use these features. If it is the case that most of these features are needed the interface to SketchBio will eventually become as complex as the sum of the tools it wraps. Because our collaborators have been able to develop models and animations without using most of the tools, and because most of the rendering settings are not needed for molecular animations, we do not anticipate this happening in practice.

## Conclusions

SketchBio is a new tool that enables scientists to rapidly construct and validate hypothetical macromolecular structures, to animate these structures, and to produce high-quality rendered animations. It has been tested and shown to meet its design goals:

- **Easy to learn and to use.** Scientists were able to rapidly construct models and animations on their own.
- **Support molecular operations.** By harnessing Pymol and Chimera.
- **Appropriately constrain layout.** Configurable collision detection, fixed-length springs, and crystal-by-example support all listed cases.
- **Support rapidly iterated, in-context design.** Real-time 6-degree-of-freedom interaction, live animation preview, and viewpoint control enable embedded design.
- **Support high-quality rendering.** By harnessing Blender.

SketchBio includes state-of-the art bimanual interaction, drop shadows to improve depth perception, and other standard modeling and animation behaviors (grouping, spline interpolation, level-of-detail rendering, rapid collision detection, real-time preview).

SketchBio also includes novel interaction and computational techniques that directly support the construction of macromolecular structures. Crystal by example and pose-mode physics both provide improved modeling capabilities and both enable more-rapid collision detection. Spring connectors show unspecified interactions and support semi-automatic structure formation. These novel capabilities can be added to existing molecular modeling tools and included in future tools, providing the same acceleration of model building and evaluation.

Both crystal-by-example and pose-mode physics enable real-time collision-detection to scale to much larger collections of molecules than are possible using existing techniques that must check for collisions among all objects. The ability to load arbitrary geometry files enables the tool to scale beyond molecule types that can be found in the protein data bank.

The design decisions (a direct-manipulation, real-time interface; harnessing tools rather than re-implementing techniques; and making a system usable in the scientists' labs) led to a system that met all of the design goals and is being used in our collaborators' own labs. The relative benefits of these decisions outweighed there potential pitfalls, making them likely choices for other designers.

SketchBio is built using portable libraries and has been compiled and used on Windows, Mac OS X, and Ubuntu Linux. Descriptions and videos of SketchBio can be found at http://sketchbio.org. SketchBio is being released as open-source software and will be available for download at this location.

## Acknowledgements

This work was supported by NIH 5-P41-EB002025.

Molecular graphics and analyses were performed with the UCSF Chimera package. Chimera is developed by the Resource for Biocomputing, Visualization, and Informatics at the University of California, San Francisco (supported by NIGMS P41-GM103311).

3D rendering was performed using Blender (blender.org). Blender is an open source project supported by the Blender Foundation and the online community.

Early versions of SketchBio used PyMOL (pymol.org) to import data from the PDB. PyMOL is an open source project maintained and distributed by Schrödinger.

We thank Ping-Lin ("Joe") Hsiao for evaluating the effectiveness of SketchBio on the construction of protofibrils. We also thank our collaborators Susan Lord, Sharon Campbell, and Kerry Bloom and the rest of the UNC CISMM group.

**List of abbreviations used**
PQP: *Proximity Query Package*, VRPN: *Virtual Reality Peripheral Network*, PDB: *protein data bank*.

**Competing interests**
The authors declare that they have no competing interests.

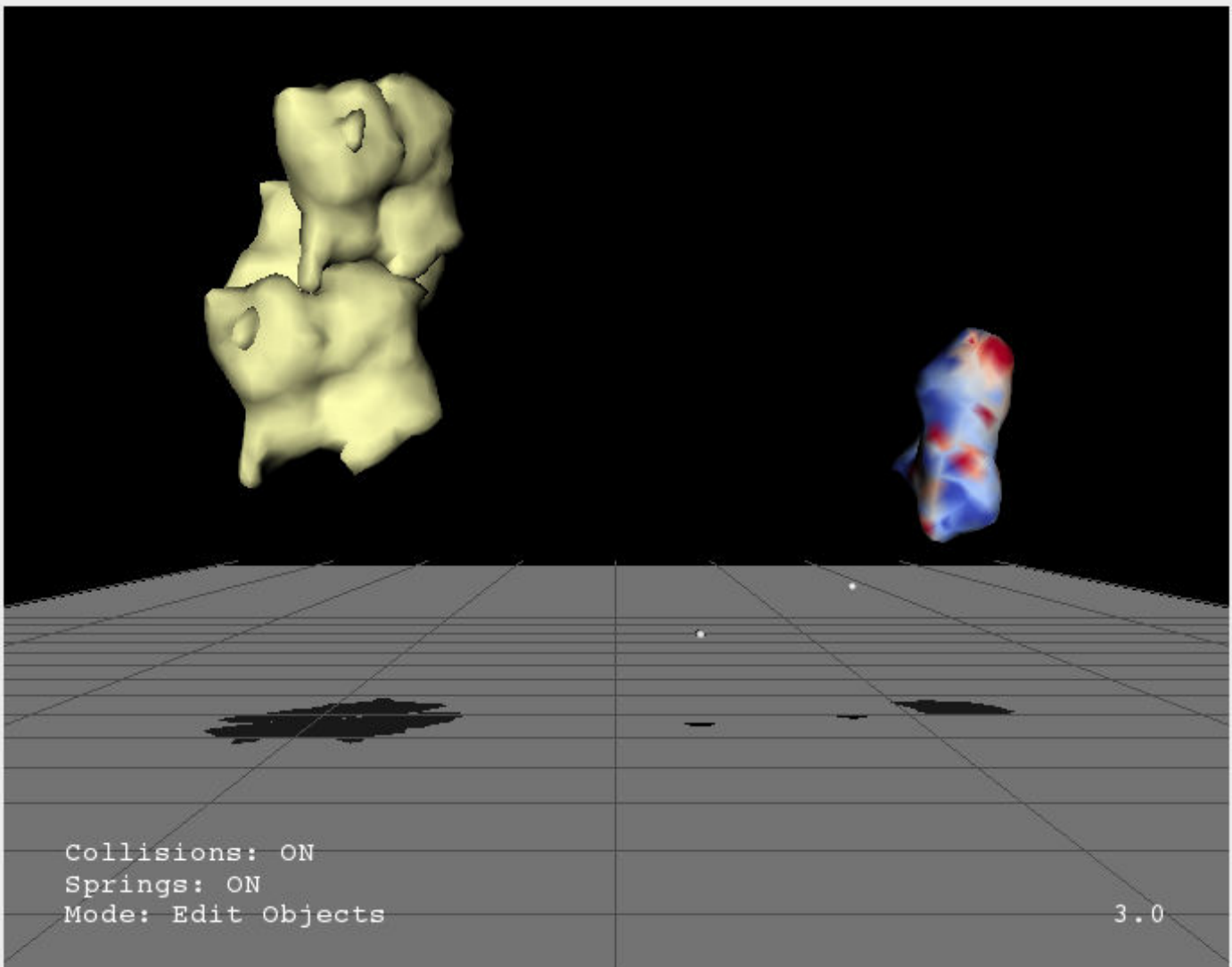

Figure 1

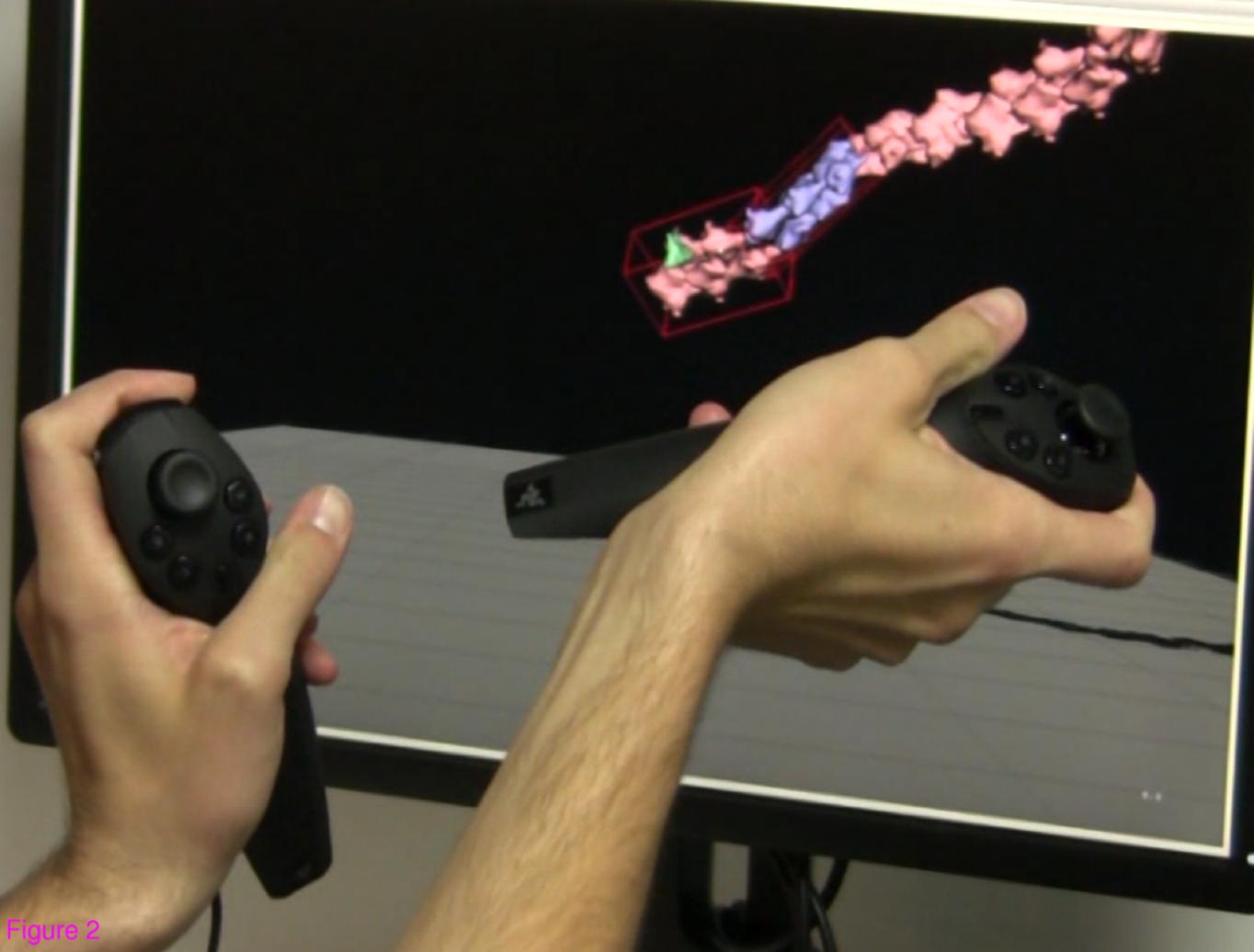

Figure 2

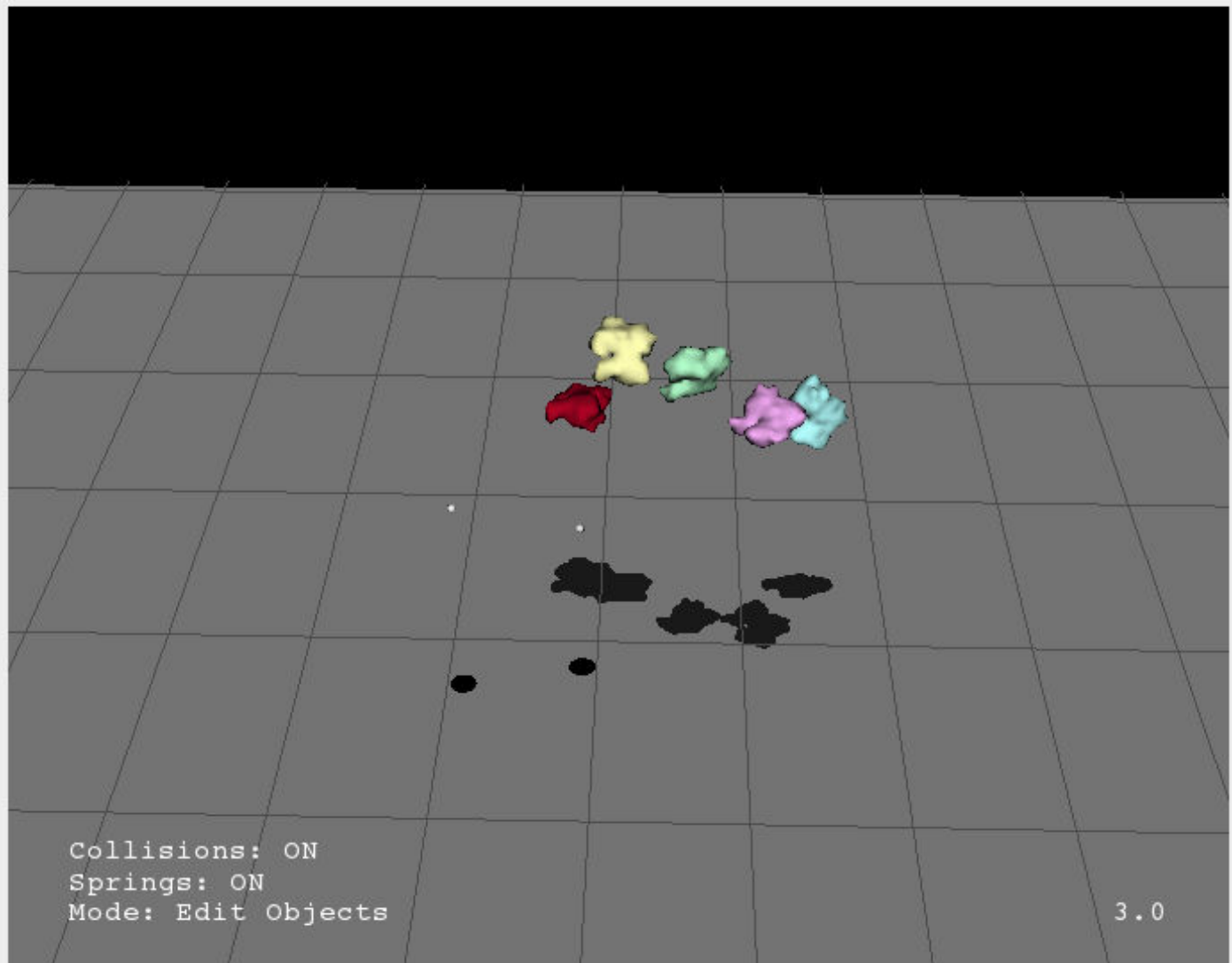

Figure 3

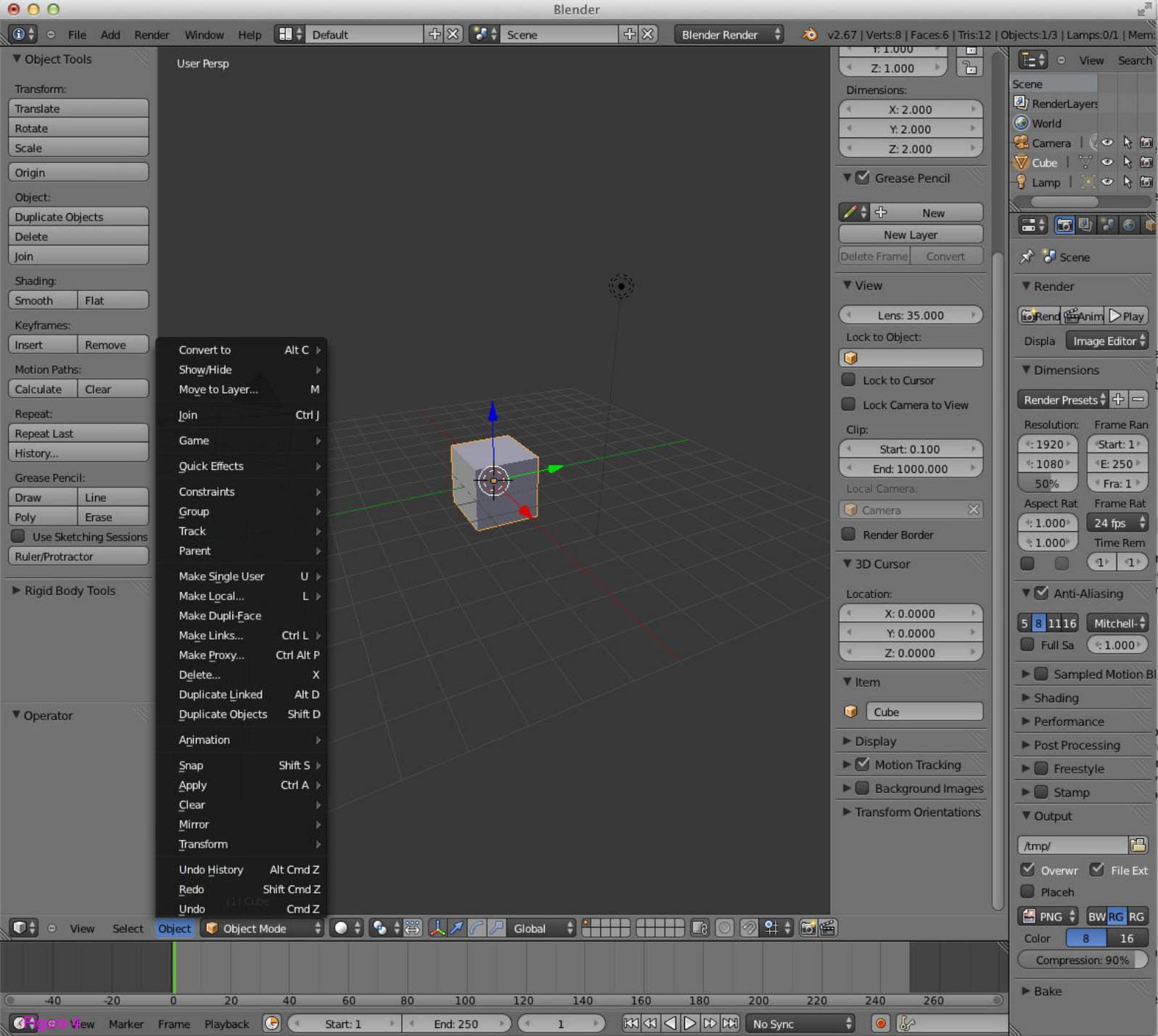

Figure 4

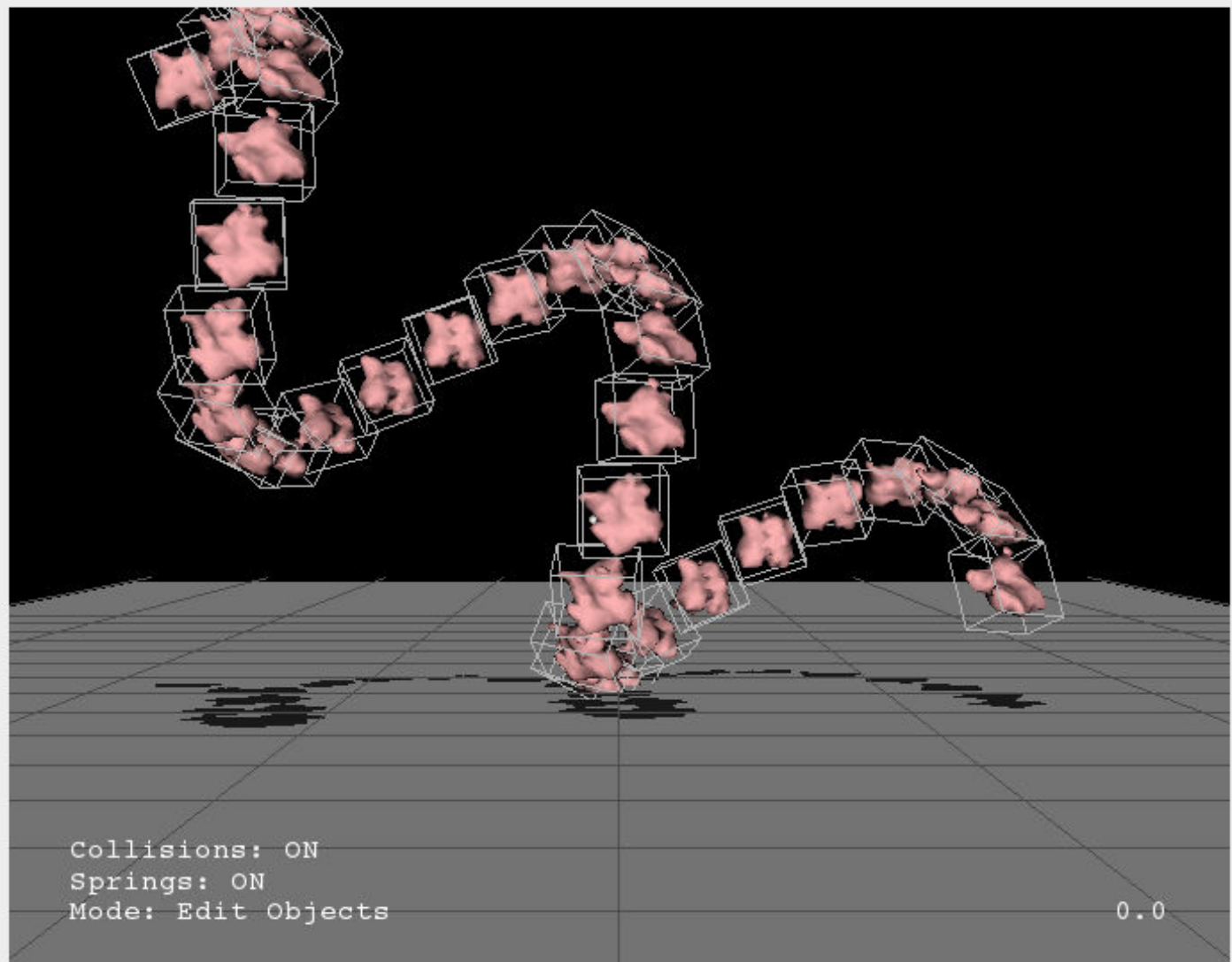

Figure 5

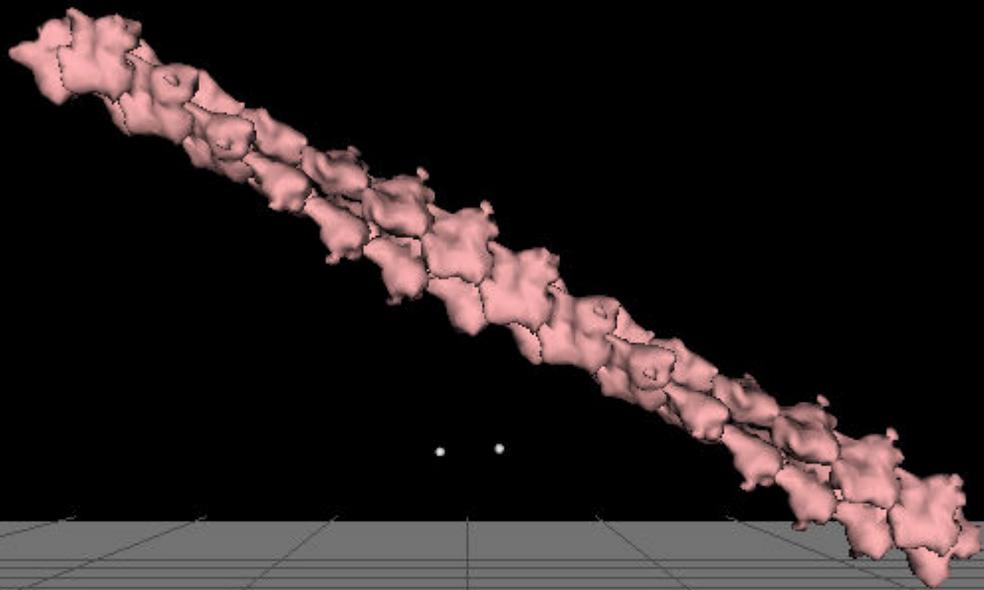

Figure 6

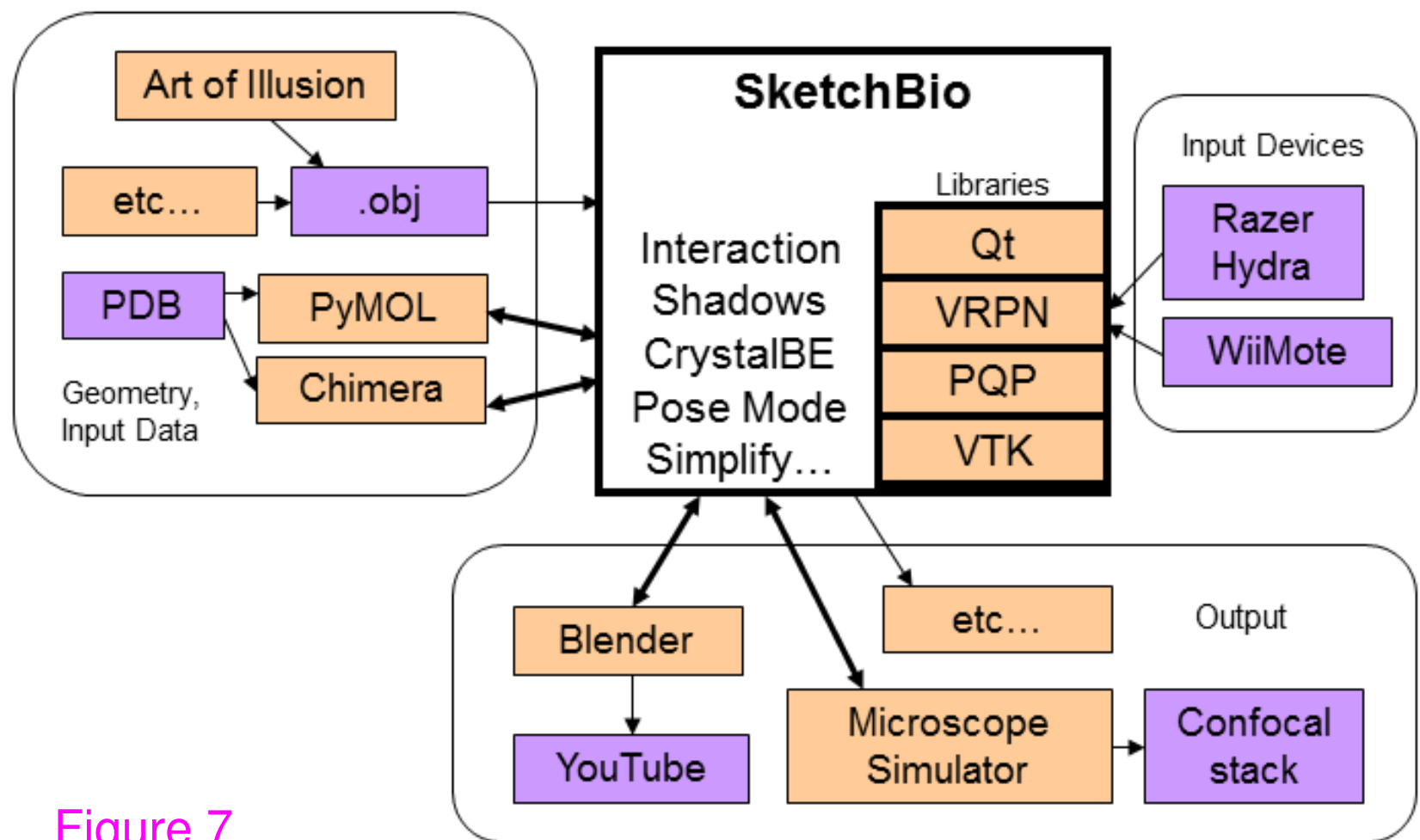

Figure 7

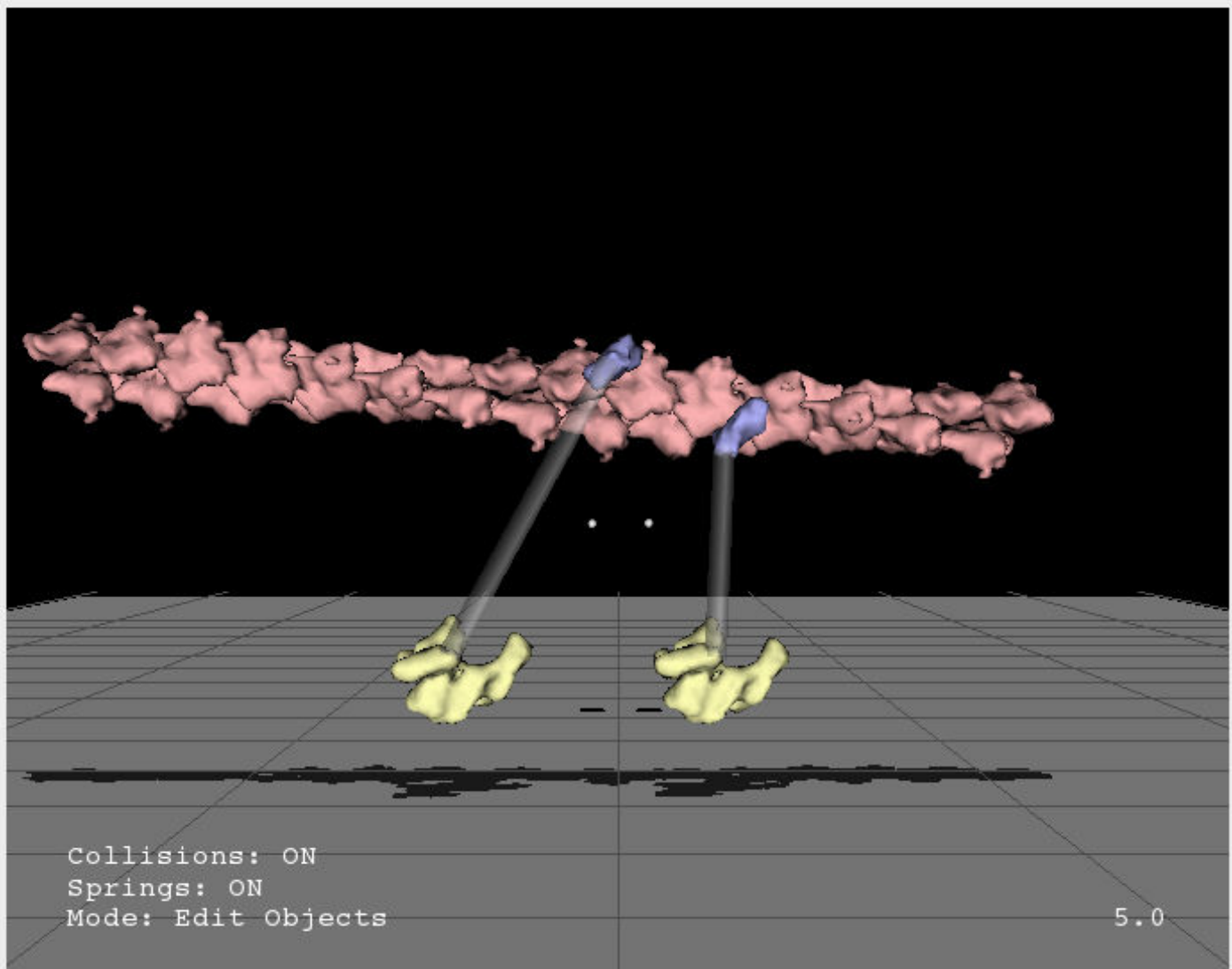

Figure 8

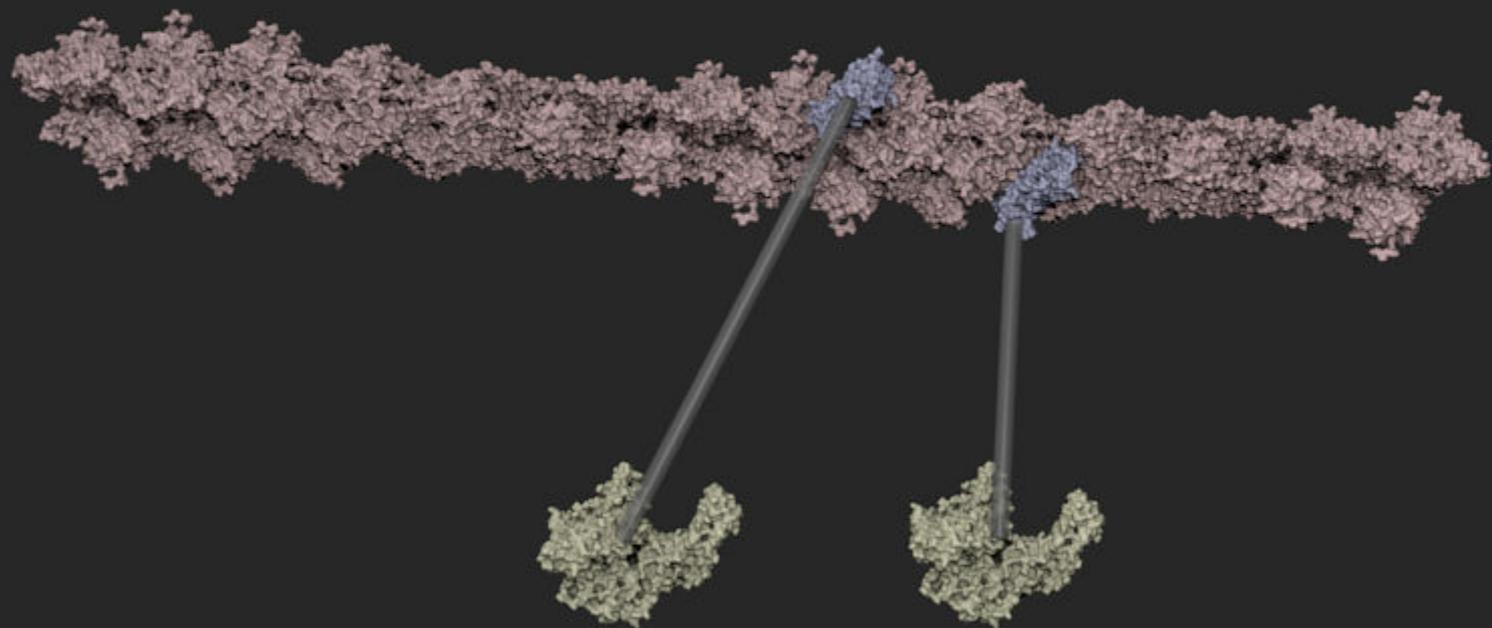
Figure 9

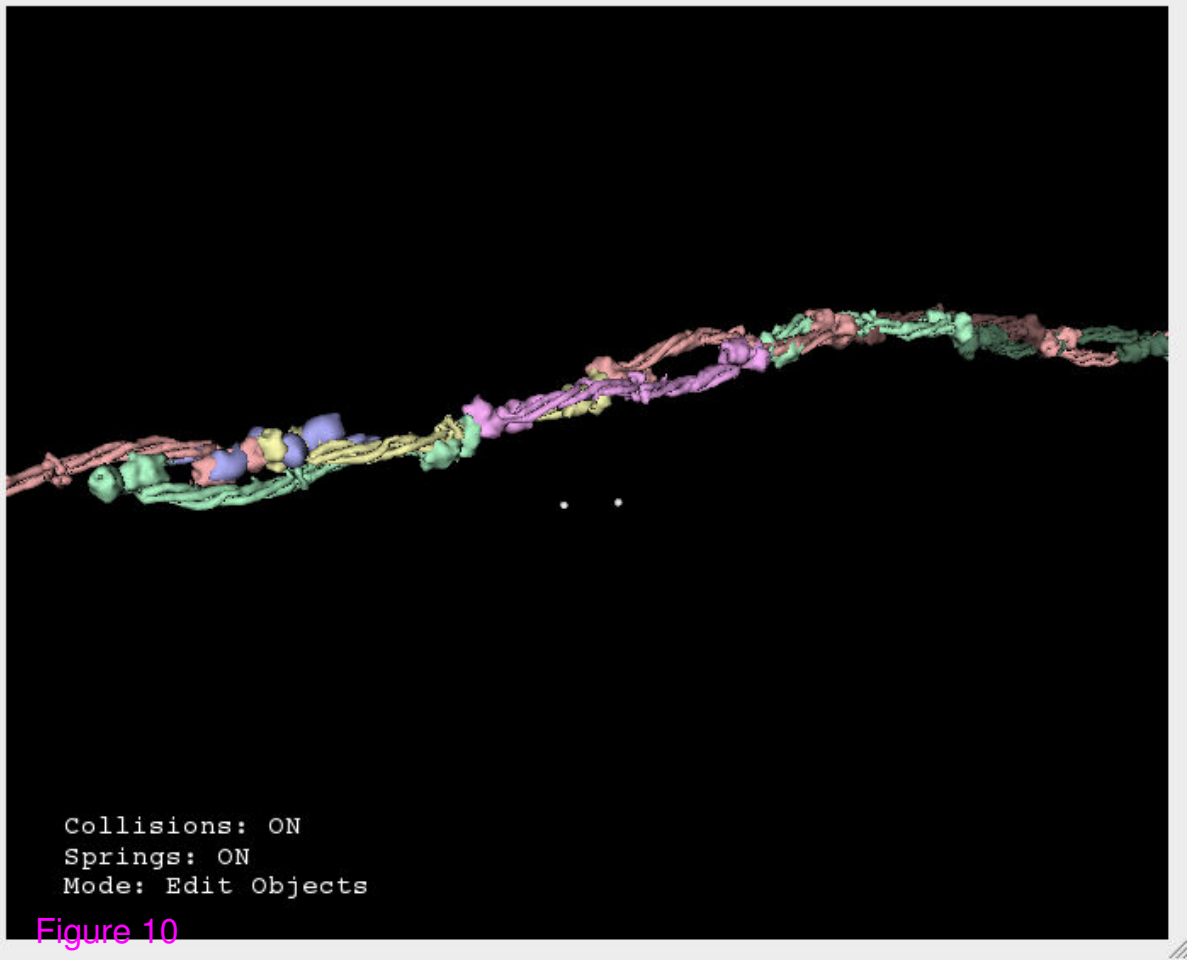

Figure 10

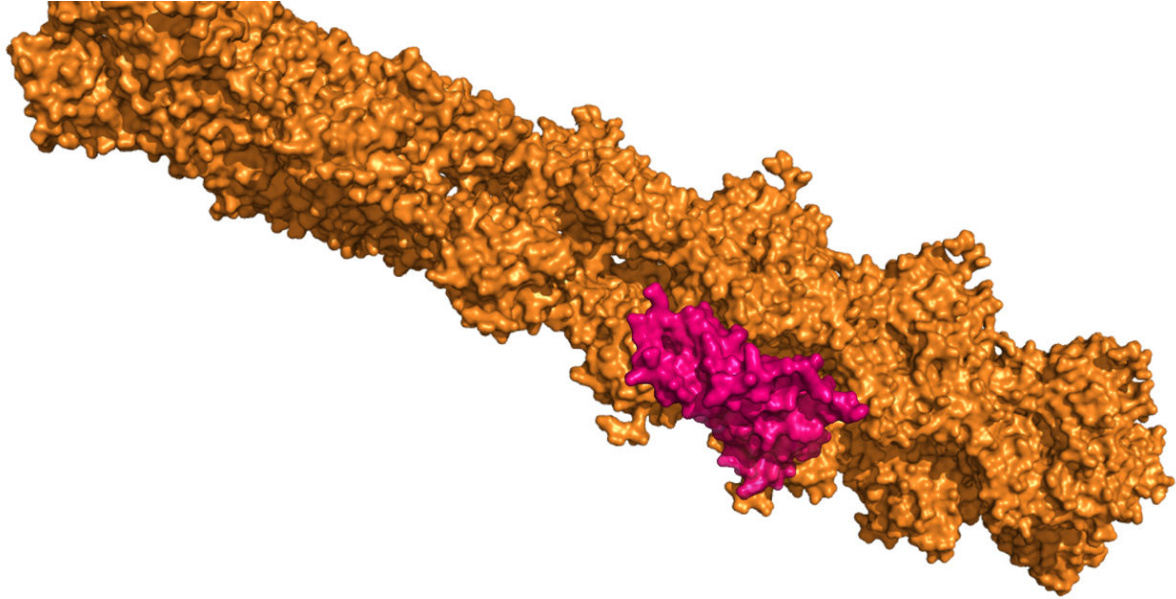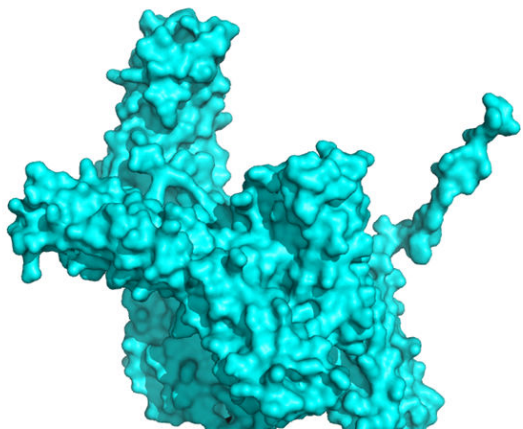

Figure 11

**Additional files provided with this submission:**

Additional file 1: sketchbio_v2_smaller.mp4, 19683K
http://www.biomedcentral.com/imedia/8189175621333842/supp1.mp4
Additional file 2: bmc_article_sketchbio.tex, 68K
http://www.biomedcentral.com/imedia/7428266221333846/supp2.tex
Additional file 3: bmc_article_sketchbio.bbl, 18K
http://www.biomedcentral.com/imedia/7254593421333848/supp3.bbl
Additional file 4: sources.bib, 15K
http://www.biomedcentral.com/imedia/1666243651133384/supp4.bib